# Programmable photonic integrated meshes for modular generation of optical entanglement links

Mark Dong,[1,2,7] Matthew Zimmermann,[1] David Heim,[1] Hyeongrak Choi,[2] Genevieve Clark,[1,2] Andrew J. Leenheer,[3] Kevin J. Palm,[1] Alex Witte,[1] Daniel Dominguez,[3] Gerald Gilbert,[4,8] Matt Eichenfield,[3,5,9] and Dirk Englund[2,6,10]

[1]*The MITRE Corporation, 202 Burlington Road, Bedford, Massachusetts 01730, USA*
[2]*Research Laboratory of Electronics, Massachusetts Institute of Technology, Cambridge, Massachusetts 02139, USA*
[3]*Sandia National Laboratories, P.O. Box 5800 Albuquerque, New Mexico, 87185, USA*
[4]*The MITRE Corporation, 200 Forrestal Road, Princeton, New Jersey 08540, USA*
[5]*College of Optical Sciences, University of Arizona, Tucson, Arizona 85719, USA*
[6]*Brookhaven National Laboratory, 98 Rochester St, Upton, New York 11973, USA*
[7]*mdong@mitre.org*
[8]*ggilbert@mitre.org*
[9]*meichen@sandia.gov*
[10]*englund@mit.edu*

**Abstract**

Large-scale generation of quantum entanglement between individually controllable qubits is at the core of quantum computing, communications, and sensing. Modular architectures of remotely-connected quantum technologies have been proposed for a variety of physical qubits, with demonstrations reported in atomic and all-photonic systems. However, an open challenge in these architectures lies in constructing high-speed and high-fidelity reconfigurable photonic networks for optically-heralded entanglement among target qubits. Here we introduce a programmable photonic integrated circuit (PIC), realized in a piezo-actuated silicon nitride (SiN)-in-oxide CMOS-compatible process, that implements an $N$ x $N$ Mach-Zehnder mesh (MZM) capable of high-speed execution of linear optical transformations. The visible-spectrum photonic integrated mesh is programmed to generate optical connectivity on up to $N$ = 8 inputs for a range of optically-heralded entanglement protocols. In particular, we experimentally demonstrated optical connections between 16 independent pairwise mode couplings through the MZM, with optical transformation fidelities averaging 0.991 +/- 0.0063. The PIC's reconfigurable optical connectivity suffices for the production of 8-qubit resource states as building blocks of larger topological cluster states for quantum computing. Our programmable PIC platform enables the fast and scalable optical switching technology necessary for network-based quantum information processors.

## Introduction

Modular quantum architectures are an attractive approach in building large-scale quantum systems,[1–3] promising improved scalability and relaxed yield requirements on the quantum hardware. The modular concept naturally lends itself to optically-connected and remotely entangled qubit candidates, such as all-photonic[4–12] or atom [13–20] and atom-like[21–28] systems with access to photon-mediated entanglement protocols.[24,25,29–32] To reach the millions of qubits required for fault-tolerant quantum computing,[5,28,33] the systems thus require scalable programmable photonic integrated circuits (PICs)[34] for the construction of large-scale entanglement networks, which depend on the high-fidelity and high-speed manipulation of many visible-to-near-infrared optical connections. For a programmable PIC to generate these dynamic optical links, the circuit needs to implement (i) a high-speed optical routing network reconfigurable on timescales much less than the coherence time of the entangled qubits and less than the duration of each remote entanglement attempt (on the order of $\mu s$[24,28]) (ii) path-erasing beam splitters between target qubits and (iii) optical phase control of individual channels to correct for path-dependent phase errors[20,25] throughout the entire network. Each of these requirements has been partially addressed in either integrated photonics[8,10,15,35] or bulk components[9,17,18,25,36,37] for binary switching, beam splitting, and phase modulation. However, there remains an outstanding challenge to develop reconfigurable many-mode switches that satisfy the requirements (i,ii,iii) for optically-heralded quantum networks on PIC platforms.

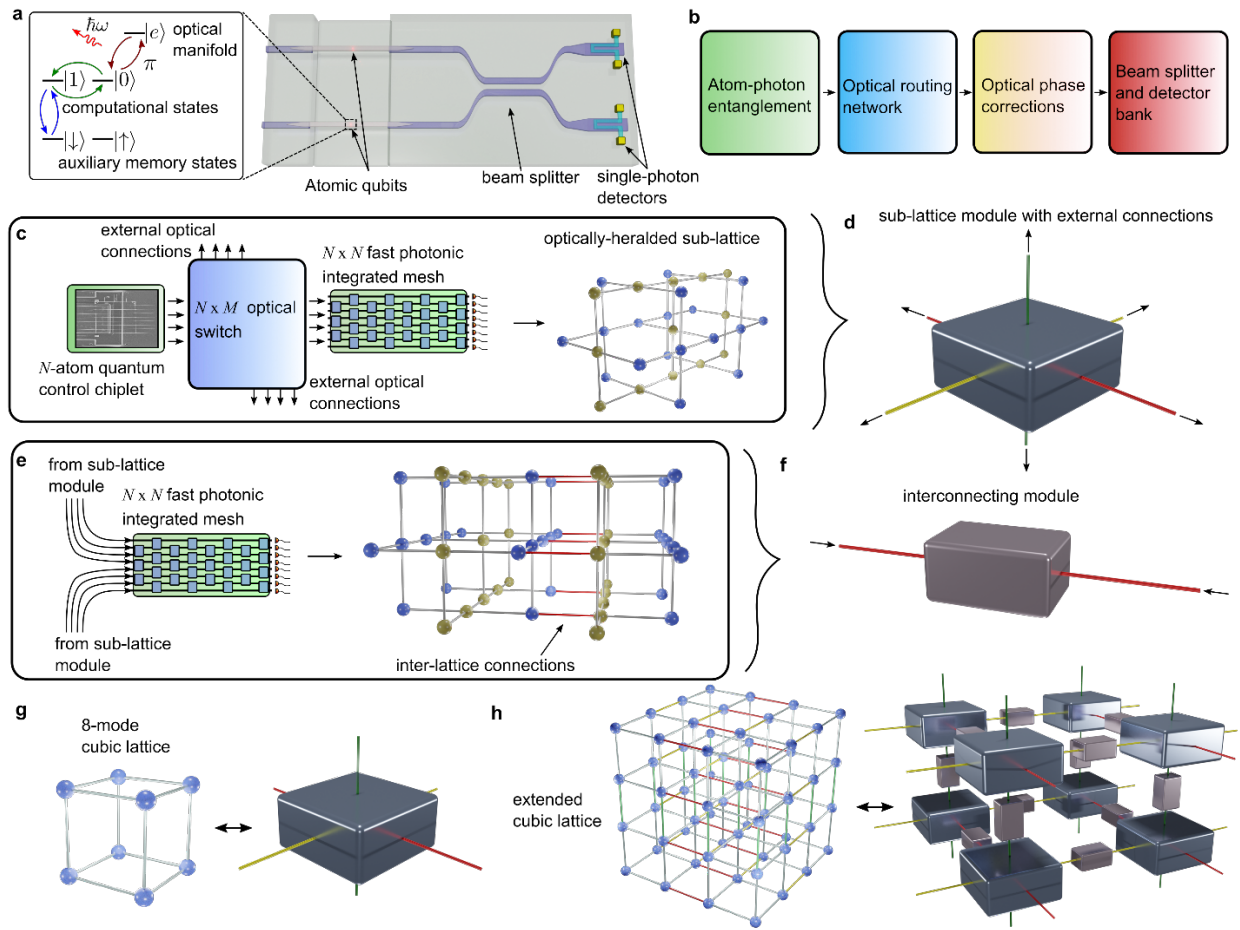

**Fig. 1: Modular architecture for generating optical links using programmable photonic integrated meshes.** a) Diagram of the basic requirements for optically-heralded entanglement generation two generic atomic qubits. The key components are initialization and control of the computational and auxiliary memory states, an optical transition emitting a photon entangled with the computational states into indistinguishable waveguide modes, a 50:50 beam splitter, and single-photon detectors. b) Schematic of the components in a) as functional blocks. c) Implementation of the block schematic using modular components: an $N$-atom qubit chiplet is connected to an $N$ x $M$ binary optical switch for photon routing, then connected to an $N$ x $N$ Mach-Zehnder mesh (MZM). The $N$ x $N$ photonic mesh performs the optical phase correction, routing to detectors, and path-erasure functions. This hardware group generates a sub-lattice of the overall quantum state. d) Render of a sub-lattice module of functions in c) showing external connections to other modules with red, green, and yellow connectors. e) Implementation of how interconnects are formed between modules using another $N$ x $N$ MZM, which links the sub-lattices. f) Render of the interconnecting module of the functions in e) that links two external connections (colored here in red). g)-h) An example of how a large-scale cubic lattice may be generated when each sub-lattice module generates a cubic unit cell.

In this work, we introduce a high-speed photonic integrated mesh with fully programmable linear-optic functions[38,39] for the modular generation of reconfigurable optical links between quantum emitters. Fig. 1 outlines our proposed architecture. The basic remote entanglement procedure involving a generic atomic qubit is shown in Fig. 1a. The qubit consists of auxiliary memory states,[40] computational states, and an optically active state coupled to only one computational state. After entanglement with the computational states, a photon is emitted and passes through a path-erasing beam splitter before detection, which heralds the entanglement in the atomic qubits. This procedure (Fig. 1b) must then be scaled and repeated[33] to generate the fully connected quantum state. Fig. 1c-f show the modularization of the remote entanglement process into two separate modules: a sub-lattice module and an interconnecting module. The sub-lattice module starts with an $N$-atom chiplet[23,41] whose optical outputs connect to an $N$ x $M$ optical switch.[36,37] The $N$ x $M$ switch then directs the $N$ emitted photons to the fast $N$ x $N$ mesh, which performs optical phase control of all inputs, additional reconfigurable optical routing, and precise 50:50 beam splitting before the detector bank, enabling the construction of an arbitrary sub-lattice of a larger quantum state (Fig. 1c-d). Alternatively, the $N$ x $M$ switch can route the $N$ emitted photons to external optical connections for the generation of inter-lattice connections using the interconnecting module (Fig. 1e-f), which contains another $N$ x $N$ mesh circuit fulfilling the same functions as those in the sub-lattice module. Fig. 1g-h shows an example of a scalable cubic lattice built from interconnected sub-lattice modules. This design architecture allows the fast PICs to configure all optical entanglement links, while the slower $N$ x $M$ switch determines which type of connection (sub-lattice or inter-lattice) is attempted. To evaluate the use of photonic meshes for generating these entanglement links, we designed and characterized a programmable PIC, fabricated on a piezo-actuated, optically broadband silicon nitride (SiN) platform[42–44] at Sandia National Laboratories. The PIC implements an 8 x 8 Mach-Zehnder mesh (MZM) based on a reversed Clements[38] design (see Supplementary section 1), allowing the execution of linear optical transformations on sub-μs time-scales. We calculated the performance of a heralded single-photon entanglement protocol[25,45] attempted on an ideal atomic qubit routed through our PIC and estimated the resulting Bell state fidelity, defined here as the optical transformation or link fidelity, which is a measure of the PIC-induced connection errors. We found the average link fidelity of 16 programmed pairwise optical connections through the 8-channel MZM exceeds 99% for the 32 possible heralded Bell states, sufficient to generate a cubic-lattice unit cell. Our results show the viability of using programmable PICs as an integral part of large-scale entanglement generation on the road towards practical quantum technologies.

**Programmable photonic integrated circuit hardware**

The schematic of our programmable 8 x 8 MZM is shown in Fig. 2. The MZM is built from two basic components (Fig. 2a): the Mach-Zehnder interferometer (MZI) with internal ($\theta_1$, $\theta_2$) and external ($\phi_1$, $\phi_2$) phase shifters, and an on-chip power monitor using a pick-off beam splitter and optical grating couplers. All optical inputs and outputs are routed from the left edge, while all electrical connections are wire-bonded from pads located at the top and bottom edges (Fig. 2c). Dummy waveguide blocks[46] are inserted where appropriate to match the optical loss and path lengths of all possible waveguide routes. The full circuit consists of 112 piezo-actuated mechanical cantilevers[44] whose DC and AC response curves (Fig. 2d and 2e respectively) show performances of $V_\pi$ ~25 V and a relatively flat frequency bandwidth up to the peak mechanical resonance of ~10 MHz. Each electrical input controls two phase shifters simultaneously ($\theta_1$-$\theta_2$ or $\phi_1$-$\phi_2$) in differential operation, resulting in two control signals per MZI for a total of 56 independent voltage channels.

We packaged and assembled the photonic and electronic system using a range of commercial and custom-built mechanical and electrical components. For optical inputs, a 20-mode optical fiber array was mounted to deliver light to all 16 input and output channels in addition to 4 calibration ports. Electrical signals were generated from two commercial boards embedded in a National Instruments PXIe system; each board has 32 channels of arbitrary waveform generation to drive the 56 electrical inputs, with each channel providing +/- 25 V. All voltages are individually programmable from a laptop using custom drivers and software interfacing with a National Instruments PXIe system. Before implementing photonic circuit configurations, we calibrated all 28 MZIs with the aid of the on-chip power monitors. See Supplementary sections 2 and 3 for more details of the electronic controls and photonic circuit calibration.

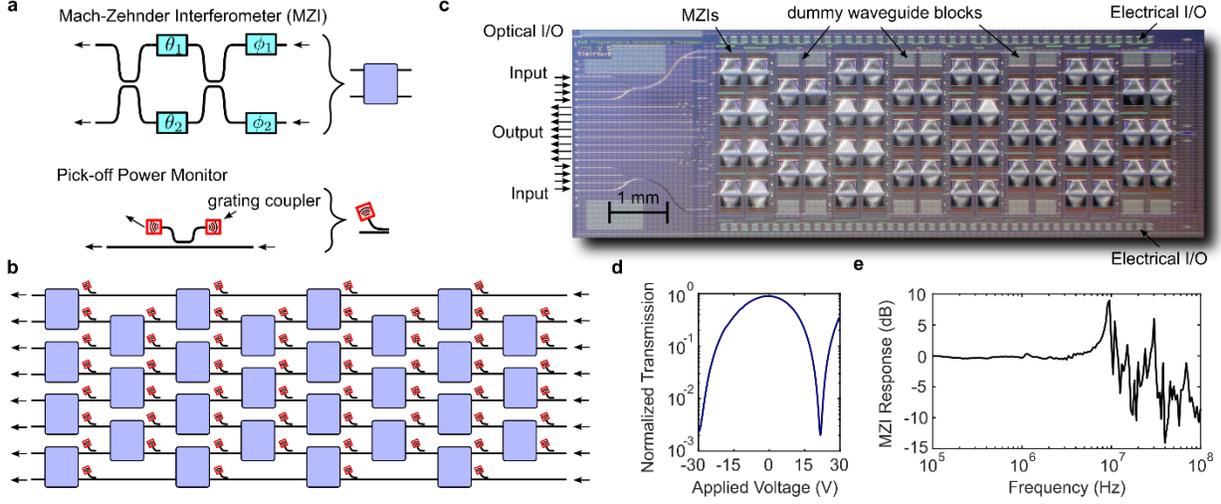

**Fig. 2: 8-channel programmable photonic Mach-Zehnder mesh.** a) Schematic of the basic building blocks of the photonic integrated mesh: a Mach-Zehnder interferometer (MZI), consisting of four optical phase-shifters and the on-chip power monitoring element, consisting of a pick-off beam splitter and grating couplers. b) Full diagram of the 8 x 8 programmable Mach-Zehnder mesh. c) Optical microscope image of the fully fabricated photonic circuit with optical and electrical I/O labeled. d) DC response when actuating phase shifts $\theta_1$-$\theta_2$ in a single Mach-Zehnder element, showing a $V_\pi \sim 25$ V. e) AC response when actuating phase shifts $\theta_1$-$\theta_2$ in a single Mach-Zehnder element, showing a relatively flat frequency response until a peak mechanical resonance at ~10 MHz.

**Quantum entanglement circuits for a cubic lattice unit cell**

We programmed the 8-mode MZM to perform the optical links for optically-heralded quantum entanglement (OHQE) as follows. Any particular OHQE circuit must fulfill the roles of optical phase control of all inputs, an optical routing network, and 50:50 beam splitters before the detector bank. Phase control of each input is achieved on-chip by the external phase shifters of the first column of the MZM. The optical routing function is implemented as part of the first seven columns of the MZM using two MZI settings: a bar state (equivalent to the identity operator $\hat{I}$), and a cross state (equivalent to the Pauli-X operator $\hat{X}$). However, because of the difficulty of reaching a perfect cross state due to fabrication errors in the passive directional couplers of each MZI,[46,47] we instead implement cross states by configuring two MZIs acting together for an error-corrected cross setting (labeled $\hat{H} - \hat{H}$).[46] The final column of the MZM is programmed as four separate Hadamard gates, labeled $\hat{H}$, which act as the path-erasing beam splitters. By configuring the columns of the MZM in this way, any of the photonic circuit's input channels can be routed to any of the output Hadamards for an entanglement operation.

We focus on a particular subset of connections out of the $\binom{8}{2} = 28$ possible connections, whose OHQE circuit diagrams (1)-(4) are shown in Fig. 3a-d respectively. Each circuit mixes four unique pairs of input channels, which results in a total of 16 possible connections for the four OHQE circuit configurations in Fig. 3. By illustrating these 16 optical links as lattice bonds and the input channels as lattice nodes (Fig. 3e), we see they are equivalent to a 3D cubic lattice unit cell with four additional (optional) connections.

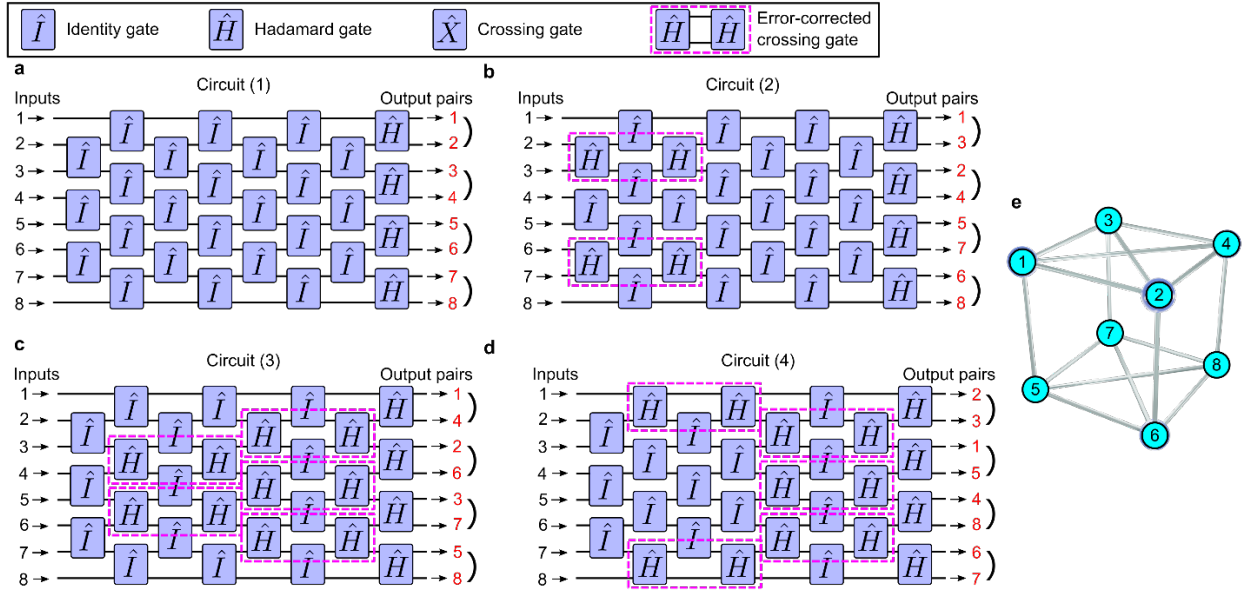

**Fig 3. Optically-heralded quantum entanglement (OHQE) circuit configurations.** a)-d) OHQE circuit diagrams labeled (1) - (4) with error-corrected crossing gates for the four circuits. Each circuit routes a pair of inputs to the output Hadamard gates, enabling photon-mediated quantum entanglement protocols. e) Connectivity generated between the 8 inputs using circuits (1)-(4), forming a cubic lattice unit cell. Connections between inputs (1,4), (2,3), (5,8), and (6,7) are optional for the purposes of generating the unit cell.

Using a coherent laser at 725 nm wavelength, we experimentally characterized the performance of the four OHQE circuits in Fig. 3 with several metrics. First, each circuit was fully programmed by setting calibrated voltages on all devices. We then characterized the optical routing efficiency by sending in one optical input at a time to collect input / output vector relations with corrections accounting for output channel grating efficiencies (see Supplementary section 4). We calculated the unitary magnitude fidelity $F^{(N)}$ performed by each circuit ($N$) using the Hilbert-Schmidt inner product[48] (Equation 1).

$$F^{(N)} = \frac{1}{8} Tr(|U^\dagger_{ideal}||U_{exp}|) \quad (1)$$

Lastly, we measured the infidelity induced by the PIC for an optically-heralded entanglement attempt by sending light in two optical inputs labeled ($i, j$), each pair corresponding to one of the four possible optical links per circuit. After balancing the input powers $I_i = I_j = I_0$ to each channel, we used the external phase shifters on the very first column of the MZM to sweep the optical phase between the two inputs, which we define as $\alpha_{ij}$. The resulting interference patterns at the corresponding output Hadamard mixing the input states ($i, j$) are given in Equations 2 and 3.[49]

$$I_n/I_0 = |u_{ni}|^2 + |u_{nj}|^2 + 2|u_{ni}||u_{nj}|\cos(\alpha_{ij}) \quad (2)$$
$$I_m/I_0 = |u_{mi}|^2 + |u_{mj}|^2 + 2|u_{mi}||u_{mj}|\cos(\alpha_{ij} + \phi_{mj}) \quad (3)$$

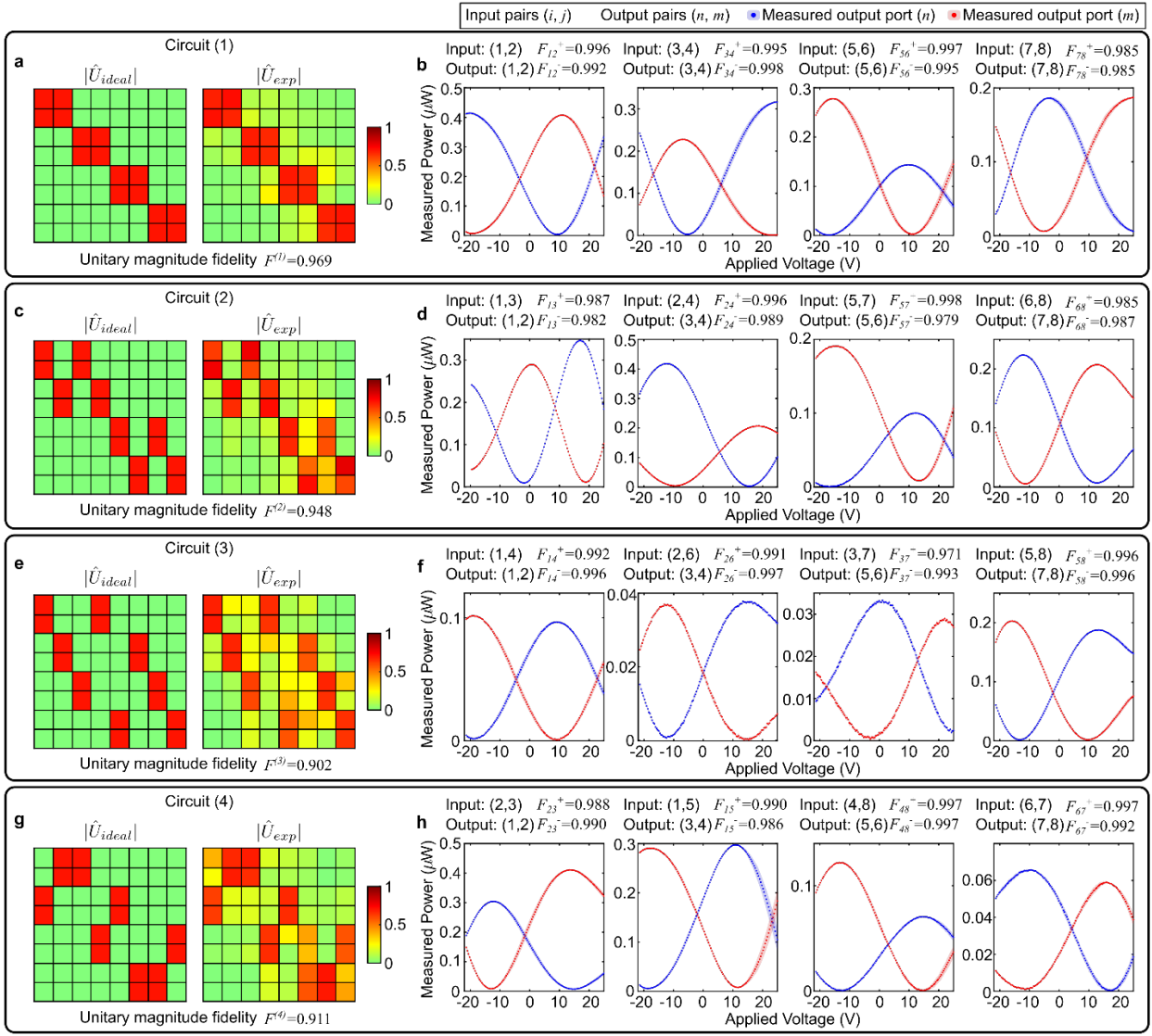

**Fig 4. Fidelities of the four programmed OHQE circuits.** a) Color map of ideal and experimental unitaries and the measured unitary fidelity of circuit (1). b) Measured intensity at the four output pairs of Hadamard operations of circuit (1) while performing an input phase sweep. Optical inputs and outputs are labeled with the convention (*A*,*B*), where *A* and *B* are port numbers. The resulting link fidelities $F_{ij}^{\pm}$ of each input pair are displayed. c)-d), e)-f), g)-h) are the measurements in a)-b) but for circuits (2), (3), and (4) respectively, demonstrating 16 independent connections. The average optical link fidelity of all 32 interference curves is 0.991 +/- 0.0063.

Here, $I_n$, $I_m$ are the optical powers of output channels (*n*,*m*) respectively and $u_{mj} = |u_{mj}|e^{i\phi_{mj}}$ is the magnitude and phase of the *m, j* unitary element. The on-chip control of $\alpha_{ij}$ compensates for phase mismatches between different channels as well as to measure the interference contrast of the outputs of each Hadamard gate. The maximum contrast is related to the optical link fidelity $F_{ij}^{\pm}$ (Equation 4), which is the heralded Bell state fidelity in the computational states after assuming perfectly indistinguishable photons emitted from identical atomic memories and no other sources of error.

$$F_{ij}^{\pm} = \left(\frac{1}{1+C_{ij}^{\pm}}\right)^{1/2} \qquad (4)$$

$C_{ij}^{\pm}$ is the interference contrast between a two input channel pair $(i, j)$ seeking to make the entanglement link, defined in Equations 5,6. We label the contrast of output channel $n$ with the superscript + (channel $m$ with superscript -) to match the notation of the heralded +/- Bell states.

$$C_{ij}^{+} = min(I_n)/max(I_n) \quad (5)$$
$$C_{ij}^{-} = min(I_m)/max(I_m) \quad (6)$$

The measured performance of the four OHQE circuits is shown in Fig. 4, which shows the ideal and measured unitary magnitudes as well as the interference patterns of the 16 optical links. The unitary magnitude fidelities (calculated from Equation 1) give a measure of the photon routing efficiency and range from 0.902 in circuit (3) to 0.969 in circuit (1). The 5-curve averages of the measured phase sweep interference patterns $I_n$ and $I_m$ are also plotted for each circuit, with error bars of +/- 1 standard deviation shown as the light shaded region. Based on the 32 interference curves, the optical link fidelities $F_{ij}^{\pm}$ range from 0.971 ($F_{37}^{+}$) to 0.998 ($F_{34}^{-}$ and $F_{57}^{+}$), with an average fidelity of 0.991 +/- 0.0063. The primary cause of routing infidelity stems from small loss imbalances from fabrication variances between the numerous optical paths, which affect the performance of both individual MZI gates and the error-corrected crossing gates – such errors should be reduced with continued improvements in the PIC fabrication process. We also note the optical link fidelities measured here likely overestimates the infidelity caused by the MZM alone, as we attribute all other minor errors in the experiment (polarization and power mismatch, phase and temperature drifts) to the PIC itself and not the surrounding experimental apparatus. Despite the non-idealities in the routing network as indicated in the unitary magnitude fidelities, the high optical link fidelities can be maintained without cross-talk between emitters by executing the entanglement protocols on the four input pairs sequentially, with a delay on the order of the emitted photon lifetime and without full circuit reconfiguration. In addition, entanglement purification using ancilla qubits[50] can improve the ~99% link-limited fidelity to local-gate limited, sub-threshold fidelity within a few steps. See Supplementary sections 2, 3, and 4 for the additional detailed derivations and experimental procedures.

**Discussion**

Advancements in reliable PIC fabrication technology, design, and packaging can further improve the viability of applying PICs to network-based quantum technologies. While current on-chip losses (-2.33 dB per depth) do not necessarily reduce entanglement fidelity,[51] losses can be improved to increase entanglement efficiency by refining the fabrication, improving PIC design[52] for the OHQE circuits of interest, or taking advantage of resonant structures.[44,53] Broadband visible wavelength detector integration,[54] scalable off-chip packaging solutions,[55] and CMOS electronic co-integration[56] would also eliminate additional bulk packaging and electronic components.

The operation of the MZM as part of the broader modular architecture confers several advantages. The photonic circuits are compatible with CMOS fabrication standards, allowing further scaling of the channel size of each module in addition to realistic mass-production of the necessary number of PICs for the full quantum computer or network. The versatility of the modular construction allows use of full connectivity and z-basis measurements for construction of other lattice shapes (see Supplementary section 5) as cluster states for measurement-based quantum computing (MBQC).[57–59] MBQC with cluster states has intrinsic advantages in handling non-deterministic entangling gates,[60] enabling not only quantum error correction codes through foliation,[61] such as surface codes with the Raussendorf lattice,[62] but also making use of the additional time dimension for a higher tolerance to errors.[60,63] In principle, our modular architecture also supports high-dimensional codes such as 3D surface codes[64] and 3D gauge color codes[65] in the MBQC framework. These codes have additional advantages such as single-shot error correction[66] and transversal non-Clifford gates by gauge fixing,[65] which can greatly reduce the resource overhead of fault-tolerant quantum computation.

**Acknowledgements**

Major funding for this work is provided by MITRE for the Quantum Moonshot Program. H.C. thanks the Shannon Fellowship and DARPA DRINQS. D.E. acknowledges partial support from Brookhaven National Laboratory, which is supported by the U.S. Department of Energy, Office of Basic Energy Sciences, under Contract No. DE-SC0012704 and the NSF RAISE TAQS program. M.E. performed this work, in part, with funding from the Center for Integrated Nanotechnologies, an Office of Science User Facility operated for the U.S. Department of Energy Office of Science. M.D. and M.Z. thank MITRE engineers L. Chan, K. Dauphinais, and S. Vergados for their support in building the photonic and electronic components. M.D. also thanks A. Menssen, A. Hermans, S. Bandyopadhyay, S. Krastanov, M. Trusheim, and R. Hamerly for useful technical discussions.

**Author contributions**

M.D., M.Z., and D.H. built the experimental setups and performed the experiments. M.Z. and M.D. designed and programmed the electronic control system. M.D. designed the photonic integrated circuit. M.E. and A.J.L., with assistance from D.D., supervised the device fabrication. H.C. and M.D., with assistance from G.G., performed the theoretical analysis. G.C., A.W., and K.P. performed additional experiments characterizing the photonic devices. M.D. and D.E. conceived the experiment and architecture. G.G., M.E., and D.E. supervised the project. M.D. wrote the manuscript with input from all authors.

**Additional information**

Supplementary information is available for experimental methods related to programming and calibrating the photonic integrated circuit.

**Competing interests**

D.E. is a scientific advisor to and holds shares in QuEra Computing.

**Data availability**

The data that support the plots within this paper are available from M.D. upon reasonable request.

# Programmable photonic integrated meshes for modular generation of optical entanglement links: Supplement / Methods

Mark Dong,[1,2,7] Matthew Zimmermann,[1] David Heim,[1] Hyeongrak Choi,[2] Genevieve Clark,[1,2] Andrew J. Leenheer,[3] Kevin J. Palm,[1] Alex Witte,[1] Daniel Dominguez,[3] Gerald Gilbert,[4,8] Matt Eichenfield,[3,5,9] and Dirk Englund,[2,6,10]

[1]*The MITRE Corporation, 202 Burlington Road, Bedford, Massachusetts 01730, USA*
[2]*Research Laboratory of Electronics, Massachusetts Institute of Technology, Cambridge, Massachusetts 02139, USA*
[3]*Sandia National Laboratories, P.O. Box 5800 Albuquerque, New Mexico, 87185, USA*
[4]*The MITRE Corporation, 200 Forrestal Road, Princeton, New Jersey 08540, USA*
[5]*College of Optical Sciences, University of Arizona, Tucson, Arizona 85719, USA*
[6]*Brookhaven National Laboratory, 98 Rochester St, Upton, New York 11973, USA*
[7]mdong@mitre.org
[8]ggilbert@mitre.org
[9]meichen@sandia.gov
[10]englund@mit.edu

## 1. Universality of reversed Clements mesh

We adopted a standard Clements design for our 8x8 MZM except with the optical inputs and outputs reversed. This design allows easy access to four MZIs at the end of the circuit to perform the important Hadamard gates on any combination of the eight inputs, which make up the core components of our entanglement circuits presented in the main text. We note that while the MZM is not strictly symmetric when reversed, we can still implement arbitrary 8x8 unitaries (up to a phase screen) by modifying the standard programming procedure to zero the upper right triangle of the target matrix. An example of programming a 4x4 unitary is shown in Fig. S1.

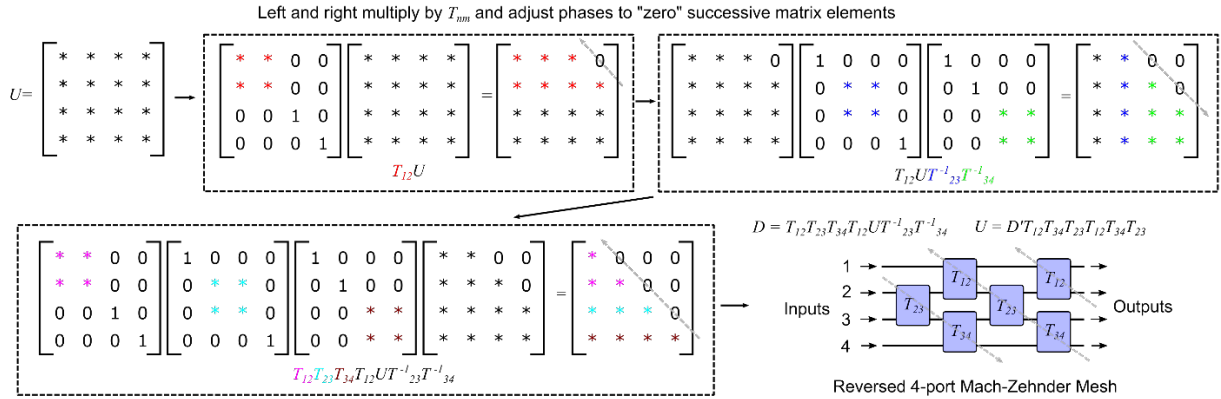

**Fig. S1: Universality of a reversed Clements mesh configuration.** The process of programming an arbitrary unitary $U$ (a 4x4 in this example) by applying successive matrices $T_{nm}$ is shown here. To generate a "reversed" Clements hardware configuration, we choose each $T_{nm}$ to "zero" a specific matrix element in the upper right triangle of $U$. The end result generates a circuit with the same number of MZIs but rotated 180 degrees, proving it is equally universal compared to the standard configuration.

## 2. Photonic integrated circuit experimental setup

Our experimental setup for characterizing and programming the 8x8 MZM is shown in Fig. S2. The photonic chip was first cleaved from the full wafer die and wire-bonded to a custom PCB. The electrical I/Os were driven by a voltage controller consisting of 64 arbitrary waveform generator channels, each with a range of +/- 25V.

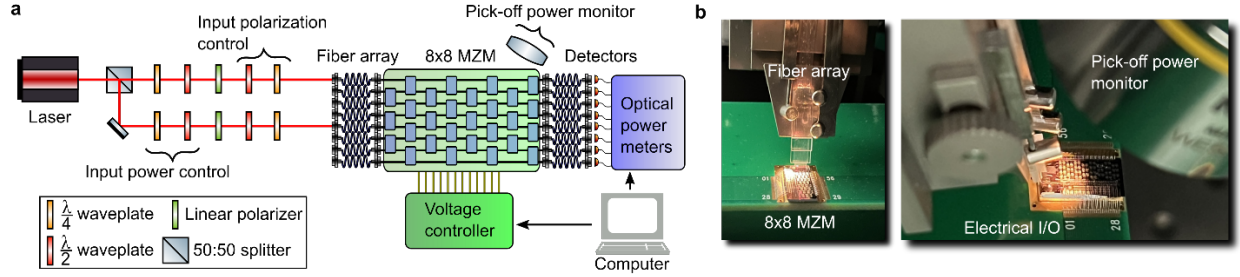

**Fig. S2: Experimental setup for 8x8 MZM.** a) Schematic of our experimental setup for characterization and programming of the 8x8 MZM. The input laser is split into two arms, each with waveplates and polarizers for independent input power and polarization control. A 20-mode fiber array feeds the input laser into the chip while simultaneously collecting output light to the detectors. The photonic chip is programmed and characterized via a laptop interfacing with a voltage controller consisting of 64, +/- 25 V channels and optical power meters consisting of 8 photodetector channels. b) Photographs of our experimental setup showing the fiber array, 8x8 MZM, electrical wirebonds, and optical microscope lens measuring the on-chip pick-off power monitors.

We used a single 725 nm wavelength tunable diode laser and split into two optical paths, each with their own power and polarization control before coupling into the 20-mode fiber array (16 fibers for the optical I/O, 4 fibers to help facilitate fiber array alignment). In order to measure all combinations of optical I/O, we reconfigured the fiber array connectors for each input pair – we only required two simultaneous optical inputs for all measurements while all optical outputs are always simultaneously collected. We monitored the power using a combination of photodiodes connected to benchtop power meters and a microscope objective measuring the pick-off grating intensity on a camera. The microscope lens and camera was mounted on a motorized stage, enabling precise access to all pick-off locations on the chip. A laptop was connected to all instruments (voltage controller, optical power meters, camera, microscope stage) as well as the programmable photonic circuit, allowing the computer to run scripts to simultaneously issue commands and collect data.

## 3. Calibration and programming procedure for 8x8 optically-heralded quantum entanglement circuits

*A. Calibration of individual MZIs*

Prior to programming any OHQE circuits, we first calibrated every MZI on our PIC. Each MZI is labeled with the convention U_#_#, where the two numbers represent the corresponding column and row of the 8x8 MZM (Fig. S3a). The procedure uses one optical input at a time and finds the best achievable bar and cross states of every MZI sequentially. The sequence depends on the specific input channel but generally attempts to isolate the optical path to only one channel of the MZI to be calibrated. To achieve the best isolation, we used a combination of diagonal optical pathing and an all bar state optical pathing. For example, when using optical input 1, the programming sequence for the diagonal path is {U_0_3, U_1_2, U_2_2, U_3_1, U_4_1, U_5_0, U_6_0} while the programming sequence for the all bar path is {U_6_0, U_4_0, U_2_0, U_0_0}. We aligned the camera and microscope to capture both output ports of each MZI and adjusted the applied voltage to find the min and max states as monitored by the pick-off grating couplers (Fig. S3b). These states were then saved as the device's cross state and bar state voltages. After cycling through all optical input ports, the circuit was fully calibrated and all bar and cross voltage values were stored in a configuration file that can be loaded for future experiments. The camera data also gave a measure of optical loss, albeit with high variance due to loss variations in each device (based on scattering intensity) as well as variations in pick-off grating coupler efficiency. We measured an average loss per depth through the chip of -2.33 dB +/- 1.87 dB, for a total loss of -16 dB to -20 dB depending on the path taken. These loss numbers include those from the pick-off structures, which sample ~ -0.5 dB per depth. Including input and output coupling, the end-to-end insertion loss of the PIC is approximately -32 to -36 dB.

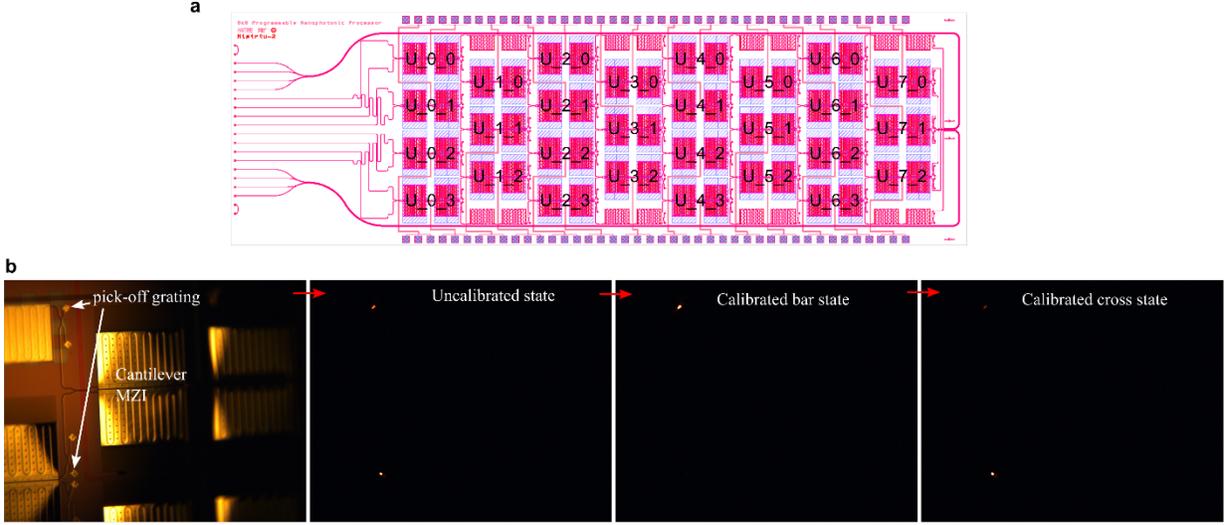

**Fig. S3: Programming and calibration procedure of the 8x8 MZM.** a) MZI device labels overlaid over the design drawing of the 8x8 MZM. The circuit is calibrated by individually characterizing all 28 MZIs using the on-chip pick-off power monitors. b) Example photos of light collected from the pick-off gratings imaged from our microscope objective. From left to right: (bright) a reference photo with backlight illumination showing features of the device under calibration; (dark) uncalibrated MZI with light going to both output ports; (dark) a calibrated bar state with light going only to the top port; (dark) a calibrated cross state with most light going to the bottom port and some leakage into the top port.

*B. Programming of double-MZI cross states*

As part of OHQE circuits (2), (3), and (4), we configured an error-corrected cross state by linking two MZIs spanning three columns of the MZM to form a double-MZI structure[46] (Fig. S4). We first set the two internal phase shifters $\theta_L$ and $\theta_R$ of the double-MZI structure (which act as tunable beam splitters) to their 50:50 state, and then adjusted the external phase shifter $\phi_R$ of the second MZI to route the light for the whole double-MZI structure. This configuration theoretically allows a perfect cross state operation, having eliminated the effect of imperfect directional couplers.

We implemented the double-MZI cross state with some optimization based on our python software scripts. We first set the two MZIs to nominally their 50:50 states. We then swept the voltage of the external phase shifter of the second MZI, monitoring the outputs to find when the bar state power is minimized. Finally, we ran a simplex optimization algorithm (Nelder-Mead algorithm) to fine-tune the voltages of the three phase shifters (two internal, one external) for the best cross state operation.[67] We saved these settings in a configuration file so they can be loaded for future experiments. We note that the experimental performance of the double-MZI cross state configured here did not always yield perfect bar port extinction. The optical paths of the two arms of the double-MZI are much longer than that of a single MZI. Moreover, the two paths traverse through two additional nearby MZIs, each set to a bar state $\hat{I}$. Both of these effects contribute to imbalanced losses in the two arms, which decreases the overall extinction of the achievable error-corrected cross state. We achieved double-MZI cross states ranging from 10 dB - 20 dB extinction due to varying local performances of the MZIs.

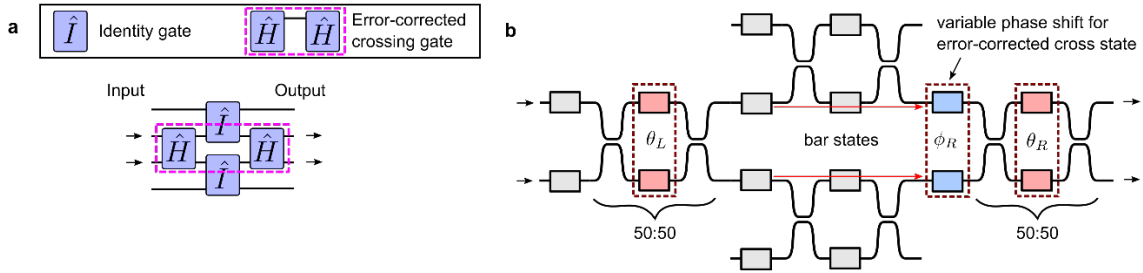

**Fig. S4: Implementation of an error-corrected crossing gate with a double-MZI configuration** a) A circuit diagram of a double-MZI implemented in the MZM. b) A detailed schematic and labels of adjusted phase shifters of the the four MZI structure representing the circuit diagram. The first MZI splits the light 50:50, then each arm has to pass through an additional MZI, each set to a bar state, before recombining at the second MZI also set to 50:50. The external phase shifter of the right MZI ($\phi_R$) is adjusted to reach the error-corrected cross state.

*C. Programming of Hadamard gates*

To accurately program the Hadamard gates in the last column of our circuits, we followed a different calibration procedure that accounts for differences in the collection efficiency and losses at each pair of Hadamard output channels. Nominally, each input $i$ or $j$ is routed to only one of the inputs of the mixing Hadamard's. By monitoring the splitting ratio $I_n/I_m$ (defined in Equations 2 and 3) when only one input ($i$ or $j$) is active, we adjusted the internal phase shifter of the MZI implementing the Hadamard gate until the output splitting ratio was precisely equal for both inputs $i$ and $j$ separately. This procedure ensures the Hadamard is calibrated to 50:50 and compensates for the different output collection efficiencies seen in Fig. 4 in the main text (due to on-chip grating variations, fiber array misalignment etc.). The final split state voltage was then saved into the configuration file.

*D. Phase sweep and interference contrast measurements*

To execute the phase sweeps for the measurement of $I_n$, $I_m$ (Equations 2, 3 in the main text), we controlled the phase difference $\alpha_{ij}$ between inputs $i$ and $j$ by sweeping the external phase shifters of the input MZIs (corresponding to columns 6 and 7 of the MZM in Fig. S3a). The exact combination of external phase shifters depends on the inputs $i$ or $j$. For instance, we swept the external phase shifters of U_6_0 and U_7_0 for inputs $(i, j) = (1,3)$, while we swept the external phase shifters of U_7_1 and U_6_3 for inputs $(i, j) = (4,8)$. The input phase shifters are always swept in opposite polarities to achieve the largest phase shift per volt.

The phase sweep was executed by the control laptop scripts. We applied a repetitive ~50 V peak-to-peak sawtooth wave at a frequency of approximately 35 Hz to the external phase shifters and recorded the outputs of the optical power meters of 125 data points per sweep. At this sweep frequency, the system was quasi-static with timing controlled by the laptop OS. We averaged five periods from each sweep, resulting in the plots shown in Fig. 4 of the main text. Sweep speeds are currently limited by the response of the photodiodes at nW sensitivities (~480 Hz).

### 4. Fidelity calculations

*A. Optical link fidelities between Bell states*

Following previously reported unitary characterization[49,68] and photon-mediated entanglement[25] work, we derive the expected Bell state fidelities when a single-photon heralding protocol is performed on our PIC and their relation to the experimentally measured interference contrast as follows. We define the two orthonormal spin states $|\uparrow_i\rangle$ as the bright state and $|\downarrow_j\rangle$ as the dark state, where $i$ and $j$ are indices of the corresponding optical input of the 8x8 MZM. After spin initialization, microwave pulse application, and photon emission, the unnormalized state of the two-spin system is

$$|\psi_{ij}\rangle = [|\uparrow_i\downarrow_j\rangle\hat{a}_i^\dagger + |\downarrow_i\uparrow_j\rangle\hat{a}_j^\dagger]\{|0\rangle\}+\ldots \quad (S1)$$

Here, $\hat{a}_i^\dagger$ is the creation operator for a photon in the input optical mode $i$ and $\{|0\rangle\}$ is the set of orthonormal vacuum Fock states for all output modes. These operators transform according to the relation

$$\hat{a}_i^\dagger = \sum_{k=1}^{8} u_{ki} \hat{b}_k^\dagger \qquad (S2)$$

where $u_{ki}$ are the unitary elements performed by the 8 x 8 circuit and $\hat{b}_k^\dagger$ is the creation operator for a photon in the output optical mode $k$. We define the indices $n$ and $m$ to represent the two optical output channels on which both inputs $i$ and $j$ are mixed. Upon detection of a single photon on channel $n$ (or $m$) and assuming no dark counts or double photon emissions), the state becomes

$$|\psi_{ij}\rangle_n = \frac{\hat{P}_n |\psi_{ij}\rangle}{\sqrt{\langle \psi_{ij}|\hat{P}_n|\psi_{ij}\rangle}} = \frac{1}{\sqrt{(|u_{ni}|^2+|u_{nj}|^2)}} [|\uparrow_i \downarrow_j\rangle u_{ni}|1_n\rangle + |\downarrow_i \uparrow_j\rangle u_{nj}|1_n\rangle] \qquad (S3)$$

where $|1_n\rangle$ is the single-photon Fock state for channel $n$ and $\hat{P}_n = |1_n\rangle\langle 1_n|$ is the projection operator. We note that the normalization factor has been applied after heralding and we assume all other terms in Equation S1 do not contribute to the projection. The target entangled spin state will be one of two possibilities,

$$|\Psi^+_{ij}\rangle = \frac{1}{\sqrt{2}} (|\uparrow_i \downarrow_j\rangle + |\downarrow_i \uparrow_j\rangle)|1_n\rangle \qquad (S4)$$

$$|\Psi^-_{ij}\rangle = \frac{1}{\sqrt{2}} (|\uparrow_i \downarrow_j\rangle - |\downarrow_i \uparrow_j\rangle)|1_m\rangle \qquad (S5)$$

depending on if detector $n$ (+ state) or $m$ (− state) fired. The fidelities of the target entangled state[1] are calculated to be

$$F_{ij}^+ = |\langle \Psi^+_{ij}|\psi_{ij}\rangle| = \left(\frac{|u_{ni}+u_{nj}|^2}{2(|u_{ni}|^2+|u_{nj}|^2)}\right)^{1/2} \qquad (S6)$$

$$F_{ij}^- = |\langle \Psi^-_{ij}|\psi_{ij}\rangle| = \left(\frac{|u_{mi}-u_{mj}|^2}{2(|u_{mi}|^2+|u_{mj}|^2)}\right)^{1/2} \qquad (S7)$$

which we define as the optical link fidelity. We characterized these fidelities by using a two input phase sweep measurement[49] on the inputs $i$ and $j$. We swept the external phase shifter on the input MZI of our circuit to adjust the phase difference between the two optical inputs, which we define as $\alpha_{ij}$. The interference pattern at the output channels of interest $n, m$ are given by the relations

$$I_n = |u_{ni}|^2 I_i + |u_{nj}|^2 I_j + 2\sqrt{I_i I_j} |u_{ni}||u_{nj}|\cos(\phi_{ni} - \phi_{nj} - \alpha_{ij}) \qquad (S8)$$

$$I_m = |u_{mi}|^2 I_i + |u_{mj}|^2 I_j + 2\sqrt{I_i I_j} |u_{mi}||u_{mj}|\cos(\phi_{mi} - \phi_{mj} - \alpha_{ij}) \qquad (S9)$$

where $I_{i,j}$ and $I_{n,m}$ are the input and output optical powers respectively and $u_{nj} = |u_{nj}|e^{i\phi_{nj}}$ is the magnitude and phase of the $n, j$ unitary element. We approximate $I_i \approx I_j = I_0$, which assumes all errors originate from the unitary elements themselves. Experimentally, this is nominally true due to our balancing the input intensity, but there may still remain small imbalances in the input powers or polarization imperfections. Thus, this approach puts an upper bound on the PIC-limited errors. For each sweep, we set the phases $\phi_{ni}, \phi_{nj}, \phi_{mi} = 0$ without loss of generality by absorbing them into our input and output basis vectors.[68] Our interference expressions now reduce to

$$I_n/I_0 = |u_{ni}|^2 + |u_{nj}|^2 + 2|u_{ni}||u_{nj}|\cos(\alpha_{ij}) \qquad (S10)$$

$$I_m/I_0 = |u_{mi}|^2 + |u_{mj}|^2 + 2|u_{mi}||u_{mj}|\cos(\alpha_{ij} + \phi_{mj}) \qquad (S11)$$

We define the interference contrasts as

$$C_{ij}^+ = min(I_n)/max(I_n) = \frac{(|u_{ni}|-|u_{nj}|)^2}{(|u_{ni}|+|u_{nj}|)^2} \quad (S12)$$

$$C_{ij}^- = min(I_m)/max(I_m) = \frac{(|u_{mi}|-|u_{mj}|)^2}{(|u_{mi}|+|u_{mj}|)^2} \quad (S13)$$

Finally, we rewrite the optical link fidelity expressions in terms of the interference contrast after incorporating the fact the redundant phases $\phi_{ni}, \phi_{nj}, \phi_{mi}$ have been set to zero.

$$F_{ij}^+ = \frac{1}{\sqrt{2}}\left(1 + \frac{1-C_{ij}^+}{1+C_{ij}^+}\right)^{1/2} \quad (S14)$$

$$F_{ij}^- = \frac{1}{\sqrt{2}}\left(1 - \cos(\phi_{mj})\frac{1-C_{ij}^-}{1+C_{ij}^-}\right)^{1/2} \quad (S15)$$

Here we see the role of $\phi_{mj}$ (the phase of the bottom right element of each 2x2 Hadamard gate, theoretically equal to $\pi$) in determining the phase difference of the heralded Bell states. Due to the imperfect routing operations of the circuit, the unitary phase may not be exactly equal to $\pi$. However, because the phase sweep measurement fully characterizes the phase $\phi_{mj}$, this can be corrected as part of a subsequent Bloch sphere rotation on the $|\Psi^-_{ij}\rangle$ spin state.[45] Thus for the purposes of calculating the optical link fidelity upon detection on port $m$, we modify the target state $|\Psi^-_{ij}\rangle$ to include the phase of $u_{mj}$

$$|\Psi^-_{ij}\rangle = \frac{1}{\sqrt{2}}(|\uparrow_i\downarrow_j\rangle + e^{-i\phi_{mj}}|\downarrow_i\uparrow_j\rangle)|1_m\rangle \quad (S16)$$

such that the optical link fidelity now reads

$$F_{ij}^- = |\langle\Psi^-_{ij}|\psi_{ij}\rangle| = \left(\frac{(|u_{mi}|+|u_{mj}|)^2}{2(|u_{mi}|^2+|u_{mj}|^2)}\right)^{1/2} \quad (S17)$$

The final optical link fidelities are now dependent only on the interference contrast achieved in each detected channel. We rewrite Equations S17 and S14 as

$$F_{ij}^\pm = |\langle\Psi_{ij}^\pm|\psi_{ij}\rangle| = \left(\frac{1}{1+C_{ij}^\pm}\right)^{1/2} \quad (S18)$$

The parameters $C_{ij}^\pm$ are extracted from each phase sweep measurement and inserted into Equation S18 to calculate each channels' respective fidelities.

*B. Unitary fidelities*

We estimate the unitary magnitude fidelity of our circuit as follows. We combined the extracted parameters from the phase sweep measurement above with additional intensity measurements to generate the experimental 8x8 matrix $U_{exp}$. For each of the entanglement generating circuits, we sent light on only one optical I/O at a time and recorded the output vector, which represents one column of the total matrix. Next, we applied a correction factor to each output pair *n,m* to compensate for different output coupling efficiencies, defined as

$$\gamma_{nm} = max(I_n)/max(I_m) \quad (S19)$$

where the values of $I_n$, $I_m$ are taken from the previous phase scan measurement. We then normalized each channel in this measurement to the total detected power to estimate the unitary magnitudes,

$$I_{tot} = \sum_{k=1}^{8} I_k \quad (S20)$$
$$|u_{nj}| \approx \sqrt{I_n/I_{tot}} \quad (S21)$$

We repeated this procedure for all optical inputs, thus characterizing all elements in our 8x8 matrix. The unitary fidelity is then simply calculated via the Hilbert-Schmidt inner product (Equation S22).

$$F^{(N)} = \frac{1}{8} Tr(|U^{\dagger}{}_{ideal}||U_{exp}|)) \quad (S22)$$

where $U_{ideal}$ is the target unitary for entanglement circuit (*N*).

## 5. Addendum on other lattice configurations

The full *N* x *N* MZM can perform additional optical links for all *N* to *N* connections. However, the OHQE circuits must be modified to include regular crossing gates as well as error corrected double-MZIs. Fig. S5a-e shows the schematic of five alternative circuits, which, when combined with OHQE circuits (1) and (2) from the main text (Fig. 3a-b), will form the 28 possible links between the *N*=8 input channels. While these circuits show the possibility of all connections, routing efficiencies may be low due to the numerous uncorrected crossing gates proliferating circuits (4)-(7). Fig. S5f shows the connectivity diagram of all 28 connections with the number of uncorrected gates labeled for each connection.

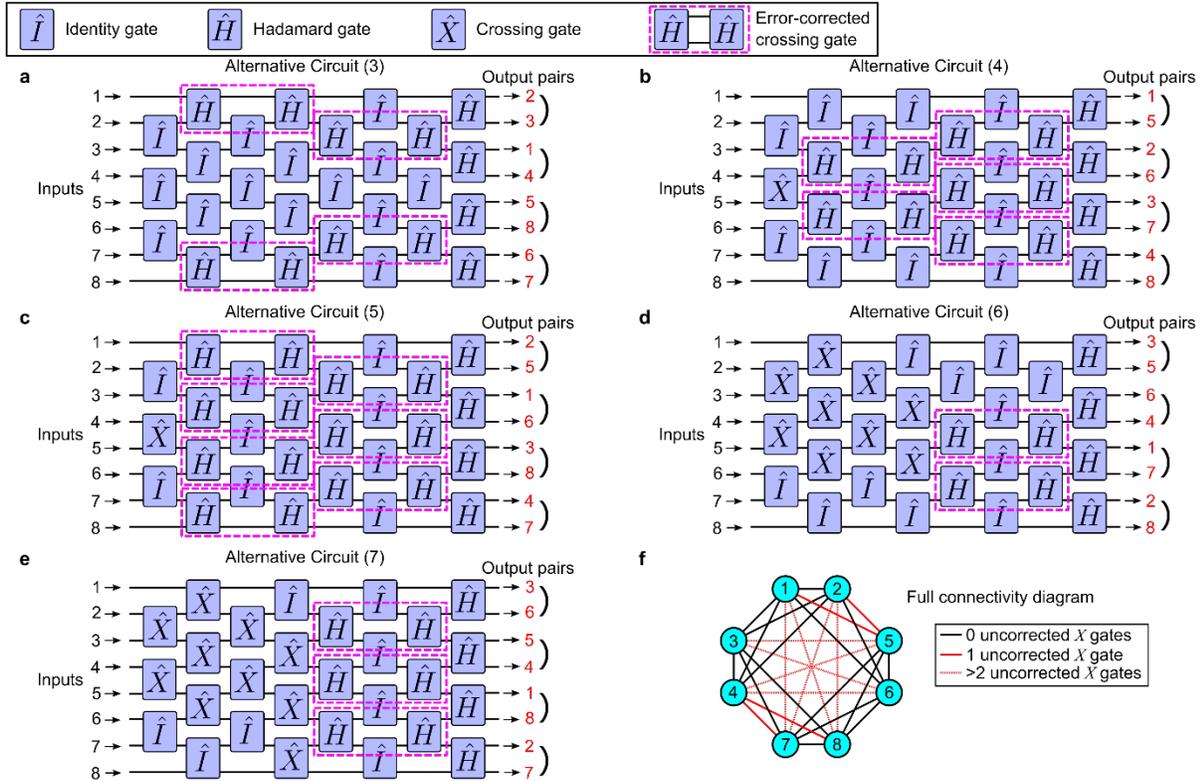

**Fig S5. Alternative circuit configurations for all-to-all connectivity.** a)-e) Alternative optically-heralded quantum entanglement circuit diagrams labeled alternative circuits (3) - (7). When combined with circuit configurations (1), (2) from the main text, these 7 circuits give all-to-all connectivity between the 8 inputs. f) Diagram of all connections between the 8 input channels and labeled number of uncorrected crossing gate operations that exist in each connection.

Full connectivity is not strictly necessary for the generation of useful 3D cluster states. We can opt to generate only cubic lattice unit cells per each submodule (Fig. 1g-h) and interconnect them to form a larger cubic lattice. From here, standard z-basis measurements on specific nodes can reduce the cubic cluster state to the topological cluster state (Raussendorf lattice) for implementation of quantum error-correcting codes. Fig. S6 shows the interconnection and measurement procedure for the construction of the 3D lattice.

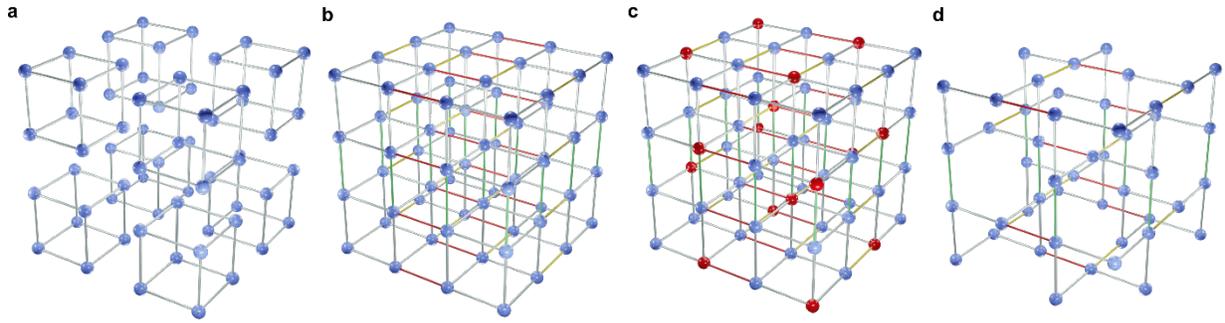

**Fig. S6: Cubic lattice construction and transformation.** a) Eight cubic lattice unit cells generated independently by individual, 8-qubit modules b) Inter-module connections are generated, shown as red, yellow, and green connections. c) Red lattice nodes are selected for a z-basis measurement in order to break specific bonds in the lattice. d) Resulting Raussendorf lattice suitable for implementation of quantum error-correcting codes.